\documentclass{article}
\usepackage{emulateapj,onecolfloat,graphics}

\begin{document}
\twocolumn
[
\title{Mock Catalogs for the DEEP2 redshift survey}
\author{Renbin Yan$^{1}$, Martin White$^{1,2}$, Alison L. Coil$^{1}$}
\affil{$^1$ Department of Astronomy, University of California,
Berkeley, CA 94720}
\affil{$^2$ Department of Physics, University of California,
Berkeley, CA 94720}

\begin{abstract}

We present a set of mock redshift catalogs, constructed from N-body
simulations, designed to mimic the DEEP2 survey.  Galaxies with a range
of luminosities are placed within virialized halos in the simulation
using a variant of the halo model.  The halo model parameters are chosen to
reproduce local clustering and abundance data and assumed to be independent
of redshift.  This allows us to predict the luminosity function, two-point
correlation function, luminosity- and scale-dependent bias, redshift space
distortions etc. of our galaxies at higher redshifts. We show that
the low order clustering properties are consistent with  preliminary
DEEP2 data. The catalogs can be used to evaluate the selection
effects of the survey and to test new algorithms and statistics that
are to be used in the analysis of DEEP2 data.
\end{abstract}

    ]

\section{Introduction}

As our theoretical understanding of structure formation has advanced
and the questions we ask of modern redshift surveys have sharpened, the use
of realistic mock catalogs has become widespread.  Almost all modern galaxy
redshift surveys have made extensive use of mock catalogs both in their
design and analysis.

Mock catalogs which reproduce the low order clustering statistics and
galaxy redshift distributions can serve to test algorithms for biases and
quantify the impact of numerous observational selection effects.
They are an indispensable tool in interpreting a wide range of observational
data.
A suitably constructed mock catalog can also allow us to connect the
observations directly to modern models of structure formation, which are
based exclusively on hierarchical build-up of dark matter halos in an
inflationary cold dark matter universe.

In this paper we describe in some detail the construction of mock
catalogs currently being used by the DEEP2 redshift survey
(Davis et al.~\cite{DEEP2}).
These mock catalogs are created by populating large, high-resolution
dark matter simulations of the $\Lambda$CDM cosmology with galaxies in
such a way as to reproduce the DEEP2 observations as closely as possible.
The mock catalogs have undergone a continual evolution over the last
few years, and we intend to continue to refine these mock catalogs as
further observations become available.  In this paper we describe the
current status of the programme and the mocks which are being used in
analysis of current DEEP2 data.
We describe some of the properties of the galaxies in the simulations
which will be useful in interpreting future DEEP2 observations and
finish by describing directions for future work.

The outline of the paper is as follows.  In \S\ref{sec:mock} we place our
work in the context of previous attempts to make mock catalogs.  The
N-body simulations and halo model are described in \S\ref{sec:sim}.
We describe the intrinsic properties of our mock galaxies, derived from
single time outputs of the simulation in \S\ref{sec:galbox} and how we
turn these `galaxy boxes' into lightcone outputs in \S\ref{sec:cone}.
After presenting some basic results in \S\ref{sec:basic} we describe
the limitations of the current generation of mock catalogs
and some directions for future development in \S\ref{sec:future}.
We conclude in \S\ref{sec:conclusions}.

\section{Mock catalogs} \label{sec:mock}

Mock catalogs have been produced in a variety of ways over the years.
Early work on mock catalogs from N-body simulations
(e.g. van Haarlem, Frenk \& White~\cite{HFW};  Cole et al.~\cite{cole};
Yoshida et al.~\cite{Yoshida}; Hamana et al.~\cite{Hamana}) 
worked with relatively low resolution simulations, which were unable to
cover a large volume while resolving the halos hosting galaxies of interest.
For this reason they made simple prescriptions for tagging individual dark
matter particles within the simulation as `galaxies'.
The probability of a DM particle becoming a galaxy was adjusted to ensure
the right mean number density of galaxies and was typically a function of
the local density, smoothed on relatively large (Mpc) scales.

Jing, Mo \& B\"{o}rner (\cite{JMB}) anticipated a halo based approach,
using what are by modern standards very small simulations.
The particles in their simulation had masses comparable to galaxy halos
and when computing correlation functions they gave such `galaxies' a
weight which was a decreasing function of halo mass.  This enabled them
to reproduce the clustering statistics of the LCRS sample
(Shectman et al.~\cite{LCRS}).
An approach along similar lines, but more akin to the procedure we adopt,
was pioneered by Kauffmann, Nusser \& Steinmetz (\cite{KNS}), 
who populated dark matter halos in a higher resolution N-body simulation with
`galaxies'.  Again individual dark matter particles were tagged as galaxies,
but the particles were now halo members and the number of halos was taken
{}from semi-analytic recipes for galaxy formation.  The key assumption of
this approach was that galaxies form in virialized dark matter halos, and that
the properties of galaxies within these halos was a function mostly
(or entirely) of the halo mass.

Mock catalogs based on a full-blown semi-analytic model of galaxy formation
were first presented in Diaferio et al.~(\cite{DKCW}), using the simulation 
described in Kauffmann et al.~(\cite{KCDW}). 
Similar work has been presented by Benson et al.~(\cite{Benson}),
Somerville \& Primack~(\cite{SP}) and Hatton et al.~(\cite{Hatton}).
Numerous mock catalogs have been produced from these highly detailed models.
Of particular relevance to the present paper is the work of
Coil et al.~(\cite{CMS}), who used semi-analytic models to create mock
catalogs for the DEEP2 survey.  The simulations presented in this paper
make a number of technical improvements over these earlier catalogs.

While the semi-analytic models have many advantages, it is difficult to
adjust them to match the known properties of any particular observational
sample, which is sometimes useful when trying to test algorithms.
What we desire is a model which situates galaxies within their correct
cosmological context, but allows us to modify their properties within
reasonable bounds to match the data more closely.

The halo model allows us to make progress in this direction.
As recognized by Peacock \& Smith (\cite{PS}), the key ingredient in the
semi-analytic modeling is the halo occupation distribution (HOD), which
describes the number of galaxies of a certain type in a halo as a function
of the host halo mass.
These authors made mock catalogs with HODs which were fitted to observations
rather than produced by semi-analytic models.
However they did not attempt to make catalogs tuned to any particular survey.
Bullock, Wechsler \& Somerville (\cite{BWS}) used the low order statistics of
Lyman Break Galaxies (LBGs; see references in Bullock et al.) to constrain
a HOD for LBGs, but again did not build any mock catalogs. 
Zhao, Jing \& B\"{o}rner (\cite{ZJB}) built mock catalogs to study the pairwise
velocity dispersions of LBGs, using a HOD formalism. 
Similar work was presented by Scoccimarro and Sheth (\cite{ScoShe}) who
used a HOD fit to the clustering seen in the PSCz
(Hamilton \& Tegmark~\cite{HT}; Szapudi et al.~\cite{Szapudi};
 Feldman et al.~\cite{Feldman})
survey to populate halos.
Their halo distribution was not based on N-body simulation but produced
using an algorithm (PT-HALOS) developed from perturbation theory. 
Mock catalogs based on high resolution N-body simulations, which resolve
the halos which host galaxies in the relevant range of luminosities, and
tuned to match both 2MASS
(White \& Kochanek ~\cite{WhiKoc}; Kochanek et al. ~\cite{2MASS})
and 2dF (Yang et al. ~\cite{YMvdBC}; Yan, Madgwick \& White 2003)
have been produced in the last few years.
This paper presents a new set for the high-$z$ galaxies probed by DEEP2.

\section{Simulations and galaxy boxes} \label{sec:sim}

\begin{table}
\begin{center}
\begin{tabular}{ccccccc}
No. & $L_{\rm box}$ & $\epsilon_{\rm soft}$ & $m_{\rm part}$ &
   $\sigma_8$ & $h$ & $n$ \\
& (Mpc/h) & (kpc/h)         & ($10^{10}M_\odot/h$) &
      &     \\ \hline
 1& 300          &    20                 &  1.7  & 1.0  & 0.70 & 1.00 \\
 2& 300          &    20                 &  1.7  & 0.8  & 0.70 & 1.00 \\
 3& 128          &     9                 &  0.1  & 0.9  & 0.70 & 0.95 \\
 4& 256          &    18                 &  1.0  & 0.9  & 0.70 & 0.95 \\
 5& 192          &    13                 &  0.4  & 0.8  & 0.67 & 1.00 \\
\end{tabular}
\end{center}
\caption{The parameters of the simulations which are being used to make
mock catalogs.  All simulations are of the $\Lambda$CDM family with
$\Omega_{\rm mat}=0.3=1-\Omega_\Lambda$.  The simulations were started
between $z=50-100$ and run to the present with output times spaced every
${\cal O}(100\,h^{-1}{\rm Mpc})$ from $z\simeq 2$.  The force softening
was of the spline form, held constant in comoving coordinates, and a
``Plummer equivalent'' smoothing is quoted in the 2nd column.}
\label{tab:nbody}
\end{table}

\subsection{N-body simulations}

The basis for the mock catalogs is a series of high resolution N-body
simulations of structure formation in a $\Lambda$CDM universe which were
run using the TreePM code (White \cite{TreePM}).
Each of the simulations has $512^3$ dark matter particles in a cubical box
with periodic boundary conditions.
The large particle number allows a wide dynamic range in mass and length
scale to be simulated.
The box sizes and other information can be found in Table \ref{tab:nbody}.
Unless otherwise stated we will draw our examples from Model 4.

We use outputs from $z\simeq 1.5$ to $z\simeq 0.6$ for DEEP2 fields with
a photo-$z$ cut.  For each output we produce a halo catalog by running a
``friends-of-friends'' group finder (e.g.~Davis et al.~\cite{DEFW}) with
a linking length $b=0.15$ (in units of the mean inter-particle spacing).
This procedure partitions the particles into equivalence classes, by linking
together all particle pairs separated by less than a distance $b$.
We keep all halos with more than 8 particles, and consider each of these
halos as a candidate for hosting `galaxies'.
The halo mass is estimated as the sum of the masses of the particles in the
FoF halo, times a small correction factor (typically 10\%) which provides
the best fit to the Sheth-Tormen mass function
(Sheth \& Tormen \cite{SheTor}).
We additionally compute a number of statistics (velocity dispersion,
virial mass etc) based on the dark matter distribution in each halo.
This information is propagated through the remaining steps in building
the mock observations, so that the halo properties and galaxy memberships
can be provided to the user if necessary.

\subsection{Galaxy populating using HOD and CLF}

\begin{table}
\begin{center}
\begin{tabular}{c|c|c|c}
No. & 2 & 3\&4 & 5 \\ \hline
$\alpha_{15}$ & -0.995 & -1.194 & -1.274	\\
$\eta$ & -0.198 & -0.280 & -0.269	\\
$M_1$  & 2.22e+11 & 1.46e+11 & 9.38e+10 \\
$M_2$  & 7.46e+11 & 1.06e+12 & 1.21e+12 \\
$\gamma_1$ & 2.649 & 1.874 & 3.612	\\
$\gamma_2$ & 0.496 & 0.366 & 0.453	\\
$\gamma_3$ & 0.656 & 0.673 & 0.622	\\
$(M/L)_0$  & 93.25 & 120.1 & 102.0	\\
$M_S$      & 7.83e+12 & 1.04e+13 & 3.86e+12
\end{tabular}
\caption{The CLF model parameters for each model we use.
For the meanings of these parameters, see Yan, Madgwick \& White (\cite{YMW})
and references therein.}
\label{tab:model}
\end{center}
\end{table}

For each output of the simulation we populate halos with `galaxies' by
choosing certain simulation particles to be galaxies and assigning them
luminosities. 
The number of galaxies more luminous than some $L_{\rm cut}$ hosted by
any halo is determined by the halo occupation distribution (HOD) which
we compute using the conditional luminosity function formulation of
Yang et al.~(\cite{YMvdB}).
The actual number of galaxies is drawn from a distribution for each halo
in the simulation which is Poisson for large galaxy numbers but drops
below Poisson for smaller numbers.
For more details, see Yan et al.~(\cite{YMW}).

\begin{figure}
\begin{center}
\resizebox{3.5in}{!}{\includegraphics{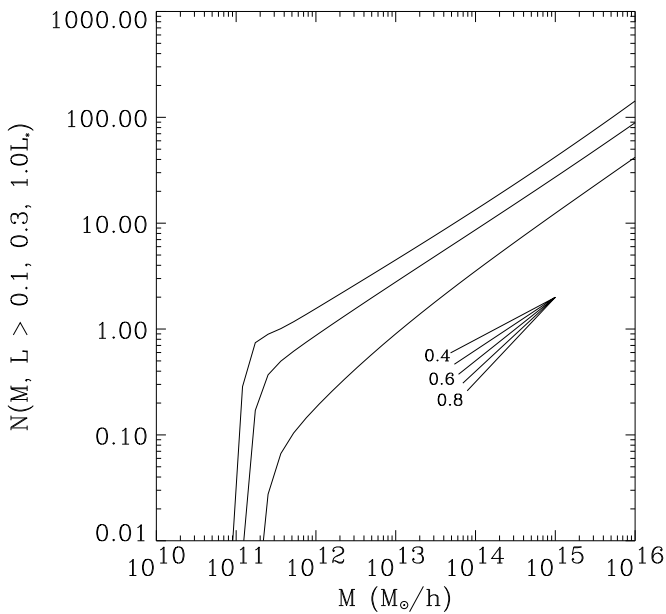}}
\caption{The halo occupation distribution we use. The three lines,
from top to bottom, are $N(M)$ for galaxies brighter than $0.1L_{*}$,
$0.3L_{*}$ and $1.0L_{*}$, respectively.  The little fan shows
several different slopes for comparison.}
\label{fig:Nm}
\end{center}
\end{figure}

Once the number of galaxies in each halo is known, they are assigned
luminosities using the conditional luminosity function (CLF) $\Phi(L|M)$
which describes the luminosity function (LF) of galaxies in halos of mass $M$.
This function is constrained from observations of the LF and clustering
statistics of the survey in question
(Yang et al.~\cite{YMvdB}; van den Bosch et al.~\cite{vdBYM};
 Yan, Madgwick \& White~\cite{YMW}).
Within the constraints set by current DEEP2 data there are a range of
cosmologies and CLFs which would be acceptable.  We further constrain
the parameter space by choosing models which fit the 2dF LF at
$z\simeq 0$ (Madgwick et al.~\cite{DSMLF}) and both the
published 2dF and lower-$z$ DEEP2
correlation functions(Madgwick et al.~\cite{mad03a} and Coil et
al.~\cite{coil}, respectively).
The method for generating the models is described in detail in
Yan, Madgwick \& White~(\cite{YMW}).
For each N-body simulation listed in Table \ref{tab:nbody} except the first,
we list in Table \ref{tab:model} one CLF model which fits the low-$z$ data
well.  For simulation 1, with a high matter clustering amplitude and a
relatively low mass resolution, we were unable to find a model which fits
the data well and which did not have a large fraction of the galaxies in
halos below our mass resolution.
To avoid multiplication of figures and tables we choose to show results for
only simulation 4 for the rest of this paper.  The main results are
unchanged for the other models.

The HODs of model 4 for three different luminosity cuts are shown in
Figure \ref{fig:Nm} and the model parameters are listed
in Table~\ref{tab:model}. 
All of the models we have chosen look qualitatively similar to the one
shown in Figure \ref{fig:Nm}: there is a sharp cut-off at low mass below
which the formation of `bright' galaxies is inefficient.  There is a
shelf or shoulder, which is required to match the number counts of galaxies,
before an approximately power-law rise to high mass.
For the models which match DEEP2 and 2dF the fits all prefer a
$\langle N_{\rm gal}\rangle$ which rises more slowly than the halo mass
itself, indicating a rising mass-to-light ratio.
This is not unexpected in samples selected in the rest-frame blue or
ultra-violet (there is some evidence that $K$-band selection provides a
steeper high end slope (Kochanek et al.~\cite{2MASS})).
That the DEEP2 galaxies are relatively underrepresented in massive halos
will have consequences for cluster finding, velocity fields and the
determination of cosmological parameters from redshift space distortions
as we shall discuss later.

Given the total number of galaxies brighter than $L_{\rm cut}$ in each halo
it is necessary to choose luminosities for them based on $\Phi(L|M)$.
Just as it was necessary to specify both $\langle N\rangle$ and the higher
moments above, it is necessary to know the fluctuations about the mean
$\Phi(L|M)$ at this stage.  If the luminosities of galaxies in each halo are
drawn independently from $\Phi(L|M)$ then one occasionally finds relatively
low mass halos with two `bright' galaxies.  Since such systems have small
radii this in turns implies an increase in the `bright' galaxy correlation
function at small scales.  Such an increase can indeed be seen in some of
the semi-analytic models, but appears to be absent in data.
This suggests that some mechanism acts to suppress pairs of bright galaxies
in small halos.  We can model this in a number of ways.  On one extreme we
could calculate the luminosities for all galaxies in halos of similar masses
by drawing from $\Phi(L|M)$ and then distribute them, round-robin, in halos
in order of decreasing luminosity.  This ensures that all the bright galaxies
are partitioned among the halos rather than having pairs end up in any one
halo.  A slightly different approach, which has very similar clustering
properties, was suggested by Yang et al.~(\cite{YMvdBC}).  Here we compute
$L_1$ such that a halo of mass $M$ has (on average) only 1 galaxy brighter
than $L_1$.  We then draw luminosities for the galaxies in this halo, allowing
only the brightest galaxy to have $L\ge L_1$.  This also suppresses the
higher moments of $\Phi(L|M)$ for bright galaxies.
We shall follow Yang et al.~(\cite{YMvdBC}) unless stated otherwise.

Once the luminosities are given, the most luminous galaxy is assigned to the
center of mass of the halo, and the other galaxies are assigned to random
particles within the halo.  
While the code allows the possibility of a radial or velocity bias in
assigning galaxies to particles, throughout we assumed that galaxies traced
the mass and velocity distribution of the halo
(inheriting its shape and any substructure).

\section{Results from the galaxy box} \label{sec:galbox}

\begin{table}
\begin{center}
\begin{tabular}{cccc}
  No.     &     2    &   3\&4      &   5   \\   \hline
$b(L_*)$  &    1.46    &   1.33        &   1.45 \\
$b'(L_*)$ &    0.22    &   0.19        &   0.22 \\
$r_0(L_*)$&    5.31    &   5.43        &   4.41 \\
$\sigma_0$&    53      &   52          &   83   \\
$\gamma$  &    0.89    &   0.80        &   0.61 \\
\end{tabular}
\caption{Some useful properties of the different models, all evaluated
at $z=1$.  The mean bias at $L_*$ and the slope of the bias with luminosity
are evaluated analytically using the CLF formulation of the halo model.
The (comoving) correlation length, $r_0$, for $L_*$ galaxies is computed from
the $z=1$ output of the N-body simulations and quoted in $h^{-1}\,$Mpc.
The last two rows correspond to the parameters in
Eq.~\protect\ref{eqn:sig_ngal}, a fit to the galaxy velocity dispersion
vs.~halo richness.
The numbers quoted here are for galaxies brighter than $0.3L_*$ and the
velocities are in (physical) km$\,{\rm s}^{-1}$.}
\label{tab:prop}
\end{center}
\end{table}

In this section, we present some results computed from these simulated
galaxy boxes.  These are not directly relevant to observations, in that
they do not include evolution, survey geometry, slitmask making, target
selection and so forth.
However they allow us to study the intrinsic properties of the mock
galaxies to set a baseline for what mock observations should later recover.

We present in Table \ref{tab:prop} some basic properties of all of our
models.
Each row of the table is discussed in corresponding sections below. 

\subsection{Luminosity Function}

\begin{figure}
\begin{center}
\resizebox{3.5in}{!}{\includegraphics{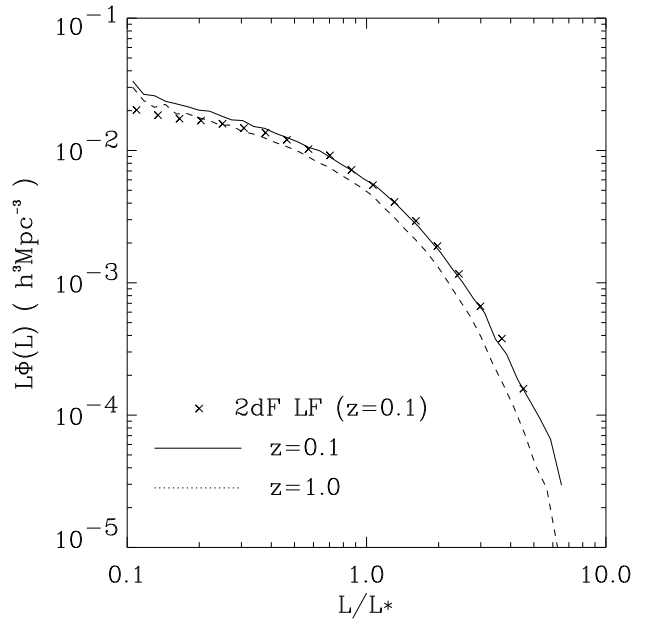}}
\caption{The luminosity functions computed directly from the galaxy boxes
at $z\simeq 0.1$ (solid line) and $z\simeq 1.0$ (dashed line).
The LF at $z\simeq 0.1$ is compared with that observed by 2dF
(Madgwick et al.~\cite{DSMLF}) (points).
Notice that the $x$-axis is $L/L_{*}(z)$, which scales out the pure
luminosity evolution.
For the LF evolution plotted in fixed magnitude bins, see
Fig.~\protect\ref{fig:LFmag}.}
\label{fig:LF}
\end{center}
\end{figure}

We first compare the luminosity function and correlation function with the
data used to constrain the model.

Figure \ref{fig:LF} shows that the simulation reproduces quite well the 2dF
LF at redshift $z=0.1$ (Madgwick et al.~\cite{DSMLF}), which we used as an
input to the model generation process.
There is a slight excess of faint galaxies in the boxes, which can be
traced to an excess of low-mass halos over the Sheth-Tormen prediction.
These excess groups are likely chance associations which are being classified
as groups by the FoF procedure at the lowest masses.
However the excess has a minimal effect on the light-cone outputs we will
generate. Due to the increasing volume and $L_{\rm cut}$ with redshift
only a small fraction of galaxies have $L<0.2L_*(z)$ in the survey.

The LF at $z=1$ has fewer relatively bright galaxies than at $z=0$ since
there are fewer massive halos.
At first sight this appears contrary to the results from the COMBO-17 survey
(Wolf et al.~\cite{COMBO-17}).  However we have plotted the luminosity in
units of $L_*$, which is a function of redshift.
As we will show later, a pure $L_*(z)$ evolution produces a LF evolution
consistent with that seen by COMBO-17.
These LFs are relatively well fit by a Schechter function, but the evolution
of the Schechter parameters is difficult to quantify because of the large
covariance between them.

\subsection{Correlation Function and Bias}

\begin{figure}
\begin{center}
\resizebox{3.5in}{!}{\includegraphics{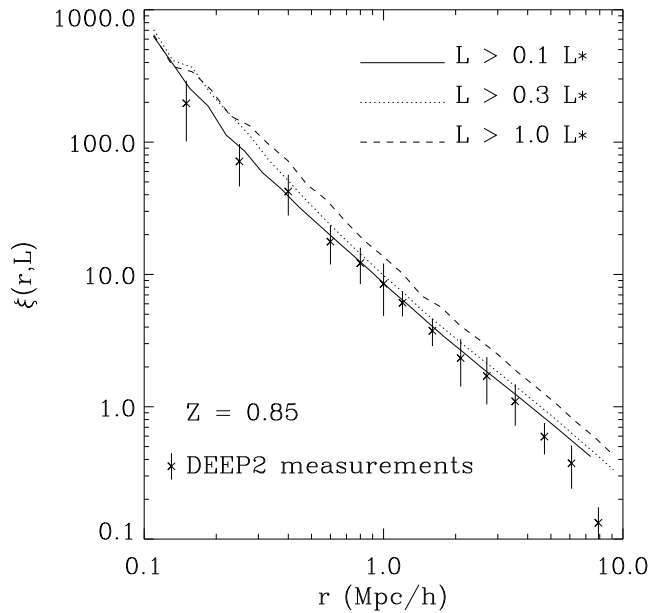}}
\caption{The correlation functions for all galaxies brighter than
$0.1L_*$, $0.3L_*$ and $1.0L_*$ in the box at redshift $z=0.85$,
compared with the DEEP2 measurements.  Here $L_*$ is $L_*(z=0.85)$.
The simulation $\xi(r)$ is computed from the periodic box, knowing the mean
density of galaxies precisely, so there are no edge effects or integral
constraint issues. The difference among the three lines shows the luminosity 
dependence of the bias.}
\label{fig:Xir}
\end{center}
\end{figure}

\begin{figure}
\begin{center}
\resizebox{3.5in}{!}{\includegraphics{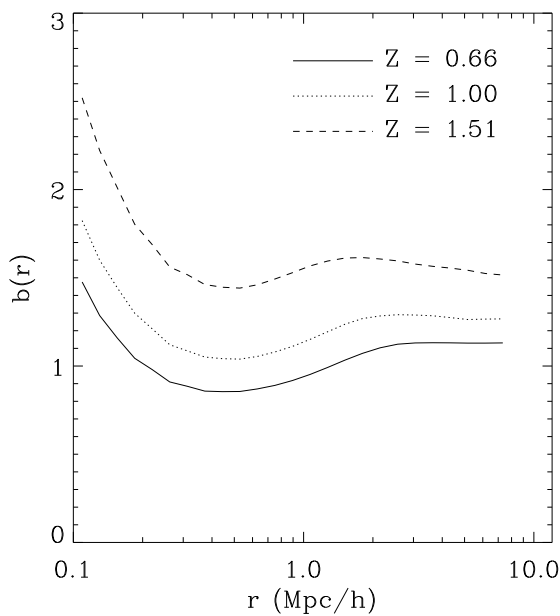}}
\caption{The scale dependent bias for several different redshifts.  The bias
is defined through the ratio of correlation functions.}
\label{fig:b_r}
\end{center}
\end{figure}

In Figure \ref{fig:Xir} we show the correlation function of all galaxies
brighter than $0.1L_{*}(z)$ in the box at $z=0.8$.  This is compared to the
data published in Coil et al.~(\cite{coil}) and again we find good agreement
\footnote{This data comes from an inversion of $w_p(r_p)$, so does not
correspond to a particular luminosity cut.  It is approximately that of a
sample of luminosity brighter than $0.1L_*$ for $L_*$ at $z=0.8$ as shown
by Fig.~\ref{fig:lumcut}.}.
The observed $\xi(r)$ is derived from the projected two point correlation
function, $w_p(r_p)$, assuming $\xi(r)$ is a power law.
This is quite a good approximation for our galaxies on the scales of
interest, and we will show later (\S\ref{sec:wprp}) that our $w_p(r_p)$ in
the mock lightcones is in good agreement with the data as expected from
this comparison.

The bias of the galaxies in our models is both luminosity and scale dependent
and stochastic, as seems to be the case for real galaxies
(Norberg et al.~\cite{Norberg1}; Norberg et al.~\cite{Norberg2};
 Peacock \cite{Peacock}; Hamilton \& Tegmark \cite{HT};
 Tegmark \& Bromley \cite{TB}; Blanton \cite{Blanton};
 Gray et al.~\cite{Gray}; Hoekstra et al.~\cite{Hoekstra}). 
In Figure \ref{fig:b_r} we show the scale dependent bias for all galaxies 
brighter than $0.1L_{*}(z)$, defined as
\begin{equation}
  b(r) = \left(\xi_{gg}(r)\over \xi_{mm}(r)\right)^{1/2} \qquad ,
\end{equation}
for several different redshifts.
In selecting the models we emphasized those in which the luminosity
dependence of the bias on large scales followed closely the measurements
made in the 2dF survey (Norberg et al.~\cite{Norberg1}) who find
\begin{equation}
  {b(L)\over b(L_*)} = 0.85 + 0.15\left( {L\over L_*} \right) \qquad .
\end{equation}
While the results are noisy due to the finite volume of the simulation
box, the power spectrum measurements bear out this expectation, though
at smaller scales there are obviously departures from this scaling.

Using the variance in counts of cubical cells we find at $z=1$ that the cross
correlation, $r$, between DM and galaxies is above 95\% on scales larger than
$8\,h^{-1}$Mpc but drops to $<60\%$ by $0.5\,h^{-1}$Mpc indicating that the
bias is stochastic.
There are insufficient galaxies in the box to probe much smaller scales but
we expect significant stochasticity in the relation between mass and galaxy
density on those scales.

At redshift $z=1$, the bias is larger than at $z=0$ in models with a fixed
HOD for all redshifts since halos of a given mass are rarer and more
biased at a higher redshift.
The luminosity-dependence of the bias is similar to that in the local
universe but with a steeper slope.
In Table \ref{tab:prop}, we show the large-scale linear bias, $b(L_*)$, and
the linear coefficient of the luminosity-bias relation around $L_*(z)$,
$b'(L_*)$, for all of our models at $z=1$.
Also shown is the correlation length, $r_0$, for $L>L_*$ galaxies in each
model at $z=1$.

\begin{figure}
\begin{center}
\resizebox{3.5in}{!}{\includegraphics{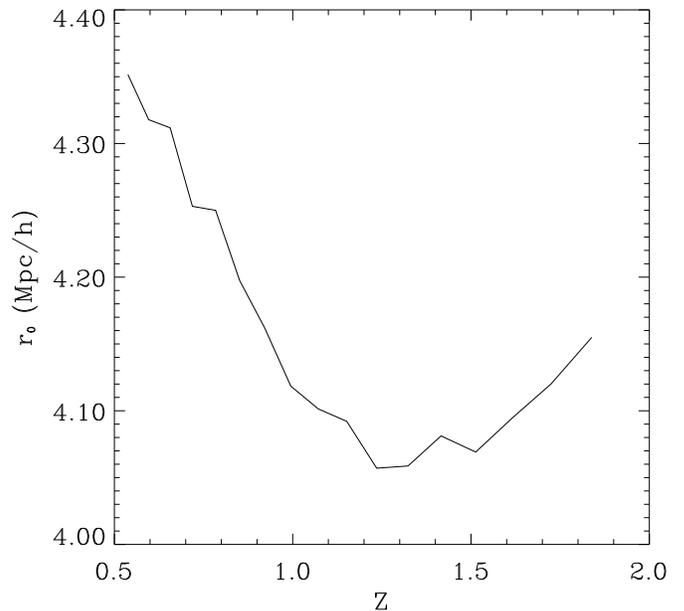}}
\caption{The redshift evolution of the (comoving) correlation length, $r_0$,
for galaxies brighter than $0.1L_*(z)$.
The observed clustering power has non-trivial evolution due to the interplay
of bias and mass clustering evolution.}
\label{fig:r0}
\end{center}
\end{figure}

The amplitude of galaxy clustering is non-monotonic with redshift, as
expected.  Figure \ref{fig:r0} shows that for galaxies brighter than 
$0.1L_*(z)$ the comoving correlation length, $r_0(z)$, has a minimum near
$z\simeq 1.2$ in these models.
Above this redshift the increasing galaxy bias dominates over the decreasing
mass clustering leading to an increase in the observed clustering strength.
Below this redshift the evolution of the bias is more modest than that of
the mass clustering.

\subsection{Redshift Space Distortion}

Of course the observations are done in redshift, rather than real, space
so it is of interest to examine the redshift space distortions in the
simulations.  We first show the distortions in the correlation function
and then look at individual halos.

\begin{figure}
\begin{center}
\resizebox{3.5in}{!}{\includegraphics{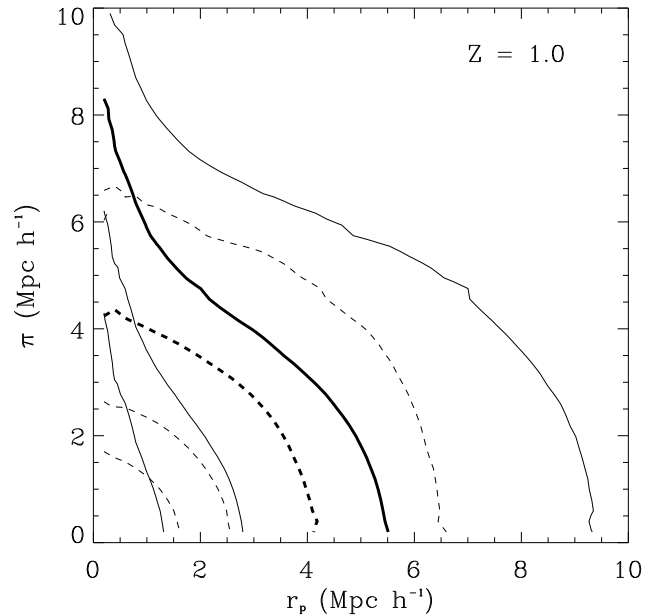}}
\caption{The 2-D correlation function in the galaxy box in real
space (dashed line) and redshift space (solid line) at $z=1.0$. The
thickest lines indicate $\xi(r_p,\pi)=1$.}
\label{fig:xisp}
\end{center}
\end{figure}

To compute the redshift space correlation function we make the assumption
that the line-of-sight is parallel to the $z$-axis of the simulation box
and define the redshift space position as $s=z+u$ where $z$ is the redshift
converted to comoving distance along the line of sight and $u$ is the velocity
converted to a line-of-sight distance difference across which the Hubble flow
is equal to the peculiar velocity.
Figure \ref{fig:xisp} shows $\xi(r_p,\pi)$ computed from the galaxy box
for galaxies brighter than $0.1L_*(z)$ at $z=1.0$.  We see the expected 
``butterfly'' structure, with large-scale
enhancement of the clustering by super-cluster in-fall (Kaiser \cite{Kaiser})
and small-scale suppression of clustering by the virial motions within halos
(the finger-of-god effect).
Transverse to the line-of-sight the clustering length is enhanced from
$r\sim 4\,h^{-1}$Mpc to $5.5\,h^{-1}$Mpc.  Along the line-of-sight the
virial motions within halos are still large enough to increase the clustering
length to $8\,h^{-1}$Mpc. 

\begin{figure}
\begin{center}
\resizebox{3.5in}{!}{\includegraphics{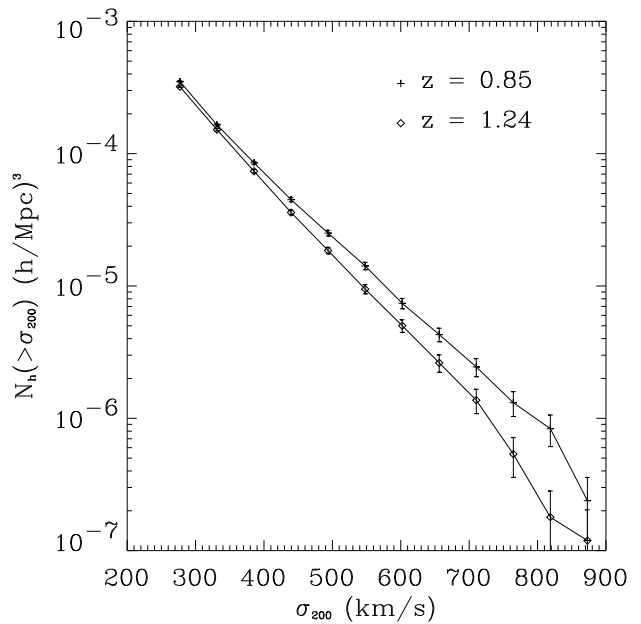}}
\caption{The cumulative (comoving) number density of halos with velocity
dispersion larger than $\sigma_{200}$, computed from the N-body simulation,
as a function of redshift.}
\label{fig:s200}
\end{center}
\end{figure}

We investigated evolution of $\xi(r_p,\pi)$, but found that the contours
show very little change over the redshift range $0.5 < z < 1.5$.
This might be in contradiction with DEEP2 observations
(Coil et al.~\cite{coil})
which show little finger-of-god effect in the higher-$z$ sample
($0.9<z<1.35$).
The source of this disagreement is currently unclear.
As we show in Figure \ref{fig:s200}, in the currently popular cosmologies
the halo population is not evolving dramatically across the redshift range
probed by DEEP2.  The reduction in the finger-of-god effect must thus be due
to the way in which galaxies are populating the dark matter halos or to an
observational selection effect.  We have assumed a constant CLF throughout,
which becomes less likely as we probe closer to the epoch at which the
majority of these galaxies are forming.  It may be that the higher redshift
galaxies are increasingly to be found outside of massive halos.  It is also
possible that at higher redshift DEEP2 is beginning to pick up a different
population of galaxies, which is significantly underrepresented in massive
halos.
On the observational side the more distant galaxies need to be brighter to
make it into the sample and their host halos subtend a smaller angle on
the sky.  Both the increased luminosity cut and the constraints imposed
during DEEP2 slitmask making could reduce the finger-of-god effect
from pairs of galaxies lying within the same halo.
We intend to investigate this, and several other issues related to
redshift space distortions, further in a future paper. 

\begin{figure}
\begin{center}
\resizebox{3.5in}{!}{\includegraphics{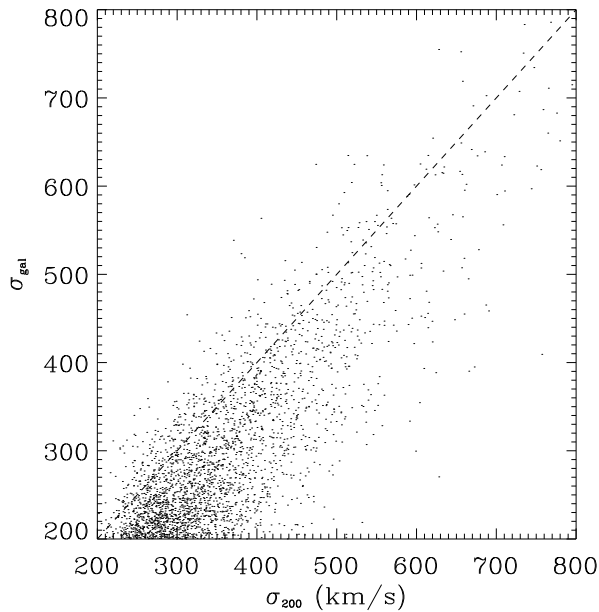}}
\caption{The comparison between the (1D) velocity dispersion of halo
member galaxies with $L>0.3L_*(z)$ and halo dark matter particles in
simulation 4 at $z=1$.
The dashed line indicates the case where the two are equal.}
\label{fig:sgals200}
\end{center}
\end{figure}

\begin{figure}
\begin{center}
\resizebox{3.5in}{!}{\includegraphics{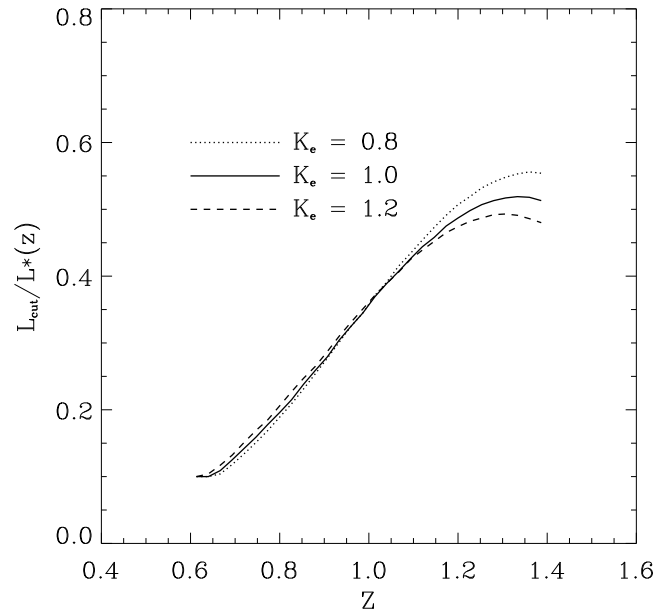}}
\caption{The minimum luminosity of DEEP2 targets as a function of $z$.
We show the minimum luminosity for 3 different assumptions for the
evolution of $M_*$, governed by the parameter $K_e$ (see \S\ref{sec:cone}).}
\label{fig:lumcut}
\end{center}
\end{figure}

We can also look at the effect of virial motions on small-scale clustering
by focusing on halo velocity dispersions.  The dark matter halos in all of
the simulations obey a reasonably tight $M-\sigma$ relation which follows
the virial scaling:
\begin{equation}
  \sigma_{200} \simeq 1100\,{\rm km}\,{\rm s}^{-1}\ E^{1/3}(z)\ 
           \left( {M_{200c}\over 10^{15}\,h^{-1}M_\odot} \right)^{1/3}
\end{equation}
where $M_{200c}$ is the mass inside a sphere interior to which the mean
density is $200\times$ the critical density, $E(z)\equiv H(z)/H_0$ is
the dimensionless Hubble parameter and $\sigma_{200}$ is the one dimensional
velocity dispersion of the dark matter particles within the same radius.
The scatter around this relation is near 5\% in $\sigma$ for $0<z<2$.
The mean richness of halos of mass $M$ can be inferred from
Figure \ref{fig:Nm} taking care to correct $M_{200c}$ to the $M_{180b}$
used in the Sheth-Tormen mass function
($M_{180b}=M_{54c}\simeq 1.4\,M_{200c}$ if $\Omega_{\rm mat}=0.3$;
 see e.g.~White \cite{HaloMass}).
We can also compute the (1D) galaxy velocity dispersion using the known halo
membership of galaxies in the box.  Even though we have taken the satellite
galaxies to have no velocity bias in these simulations, the galaxy velocity
dispersion is systematically lower than the halo velocity dispersion because
we assume that the brightest galaxy in the halo sits at the halo center and
inherits the halo center of mass velocity.
This is shown in Figure \ref{fig:sgals200}, where we plot the dispersion
computed from galaxies brighter than $0.3L_*(z)$, as appropriate for galaxies
at $z\sim 1$ (see Figure \ref{fig:lumcut}), against the halo velocity
dispersion.

\begin{figure}
\begin{center}
\resizebox{3.5in}{!}{\includegraphics{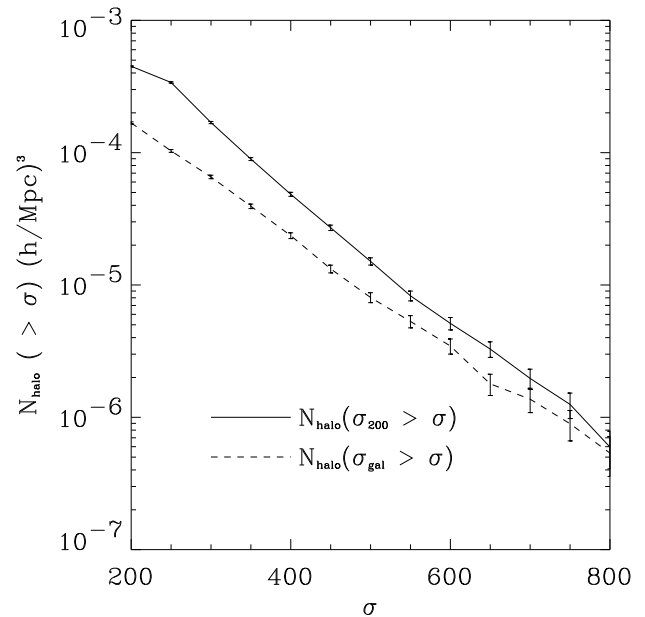}}
\caption{The cumulative (comoving) number density of halos vs.~1D velocity
dispersion in simulation 4 at $z=1$.  The solid line is the halo 
dark matter velocity dispersion
while the dashed line is for galaxies brighter than $0.3L_*(z)$.
The galaxy velocity dispersion systematically underestimates the dark
matter velocity dispersion as described in the text.}
\label{fig:Ngrpsig}
\end{center}
\end{figure}

It is also interesting from the point of view of finding groups in the
DEEP2 data to consider the cumulative (comoving) number density of groups
as a function of velocity dispersion.
This is shown in Figure \ref{fig:Ngrpsig}, at fixed redshift, again using
the known halo membership of the galaxies and only considering galaxies
brighter than $0.3L_*(z)$ at $z=1.0$.

While there is a large scatter, the richness of the groups is
correlated with the galaxy velocity dispersion.
We list in Table \ref{tab:prop} the best fitting parameters to
a power-law relation of the form
\begin{equation}
  \sigma_{\rm gal} = \sigma_0 N_{\rm gal}^{\gamma}
\label{eqn:sig_ngal}
\end{equation}
for galaxies brighter than $0.3L_*(z)$ at $z=1$ in each model we use.

\section{Survey geometry realizations and mock observation}
\label{sec:cone}

The DEEP2 redshift survey is very nearly a pencil-beam survey.
In particular the 1-hour-survey, on which we shall focus here,
will cover 4 fields each $120'\times 30'$.
For our assumed cosmology this translates into $\sim 20\times 80\,h^{-1}$Mpc
at $z=1$. 
The survey depth extends to $z\sim 1.5$ and a photo-$z$ cut will be
applied on 3 of the 4 fields to select only galaxies with $z > 0.7$.
For the remaining field no cut will be applied.
Thus for the fields with a photo-$z$ cut the mock catalog should be a
light-cone $\sim1300\,h^{-1}$Mpc (comoving) long.
For the field without a photo-$z$ cut, this would be
$\sim3000\,h^{-1}$Mpc.

Our simulation volume is 8 times as large as the survey light cone without
a photo-$z$ cut, and even larger in the case of the photo-$z$ cut.
Thus suggests we can make approximately 8 independent mock catalogs per
model from the sequence of outputs we have available. Additionally the
galaxies in earlier outputs are made independently of those in later outputs
so the effects of repeated structure are mitigated to some extent by the
galaxy population process.

To mimic the DEEP2 geometry we stack our galaxy boxes to produce light-cone
outputs.
The field of view is oriented so that the line-of-sight is offset by a few
degrees from one axis of the box.  In this way we can trace through the
structure in a continuous manner, using the periodicity of the box, without
retracing the same structure every box crossing.  We use earlier and earlier
outputs as we trace back along the line-of-sight.  In principle this can lead
to discontinuities as the structures in the box `jump' between outputs.  The
time sampling is sufficiently dense however that this is not an
issue\footnote{By design light travels ${\cal O}(100\,h^{-1}{\rm Mpc})$
between output times.  Thus a halo moving with velocity $300\,$km/s moves
only ${\cal O}(100\,h^{-1}{\rm kpc})$ between outputs.  We have tested that
this is small enough for our purposes}.

Given the distances and luminosities of the galaxies in the light-cone we
obtain magnitudes through the defining relations
\begin{equation}
  M_{B_j} = M^*_{B_j}(z) - 2.5\log {L\over L_*}
\end{equation}
\begin{equation}
  m_R = M_{B_j} -5 + 5 \log_{10} D + K(z)
\end{equation}
where $M$ is the absolute magnitude, $m$ is the apparent magnitude, $D$ is the 
luminosity distance (in pc) and $K(z)$ is the $K$-correction.
Since we used the $B_j$-band 2dF LF to tune the CLF to assign luminosities,
and DEEP2 is selecting targets in R-band, we need a $K$-correction which
transfers rest-frame $B_j$ band magnitude into observed-frame R-band
magnitude.  Our $K(z)$ is defined as 
\begin{equation}
  K(z) = R(z)- B_j(0)
\end{equation}
and we use a quadratic fit over the scattered $K$-correction points of
galaxies in DEEP2 early data from Willmer et al.~(\cite{ChrisLF}) 
to get an averaged $K$-correction over spectral types:
\begin{equation}
  K(z) = -2.85 + 5.62 z - 2.49 z^2
\end{equation}
valid for $0.6<z<1.4$.
As we will discuss later, the $y$-intercept of the $K$-correction does not
influence the redshift distribution of galaxies in the mock catalog, since
we will normalize it to the observed number counts of galaxies in the
photometric catalog. 

We write $M_*$ as a function of $z$ because it is observed that galaxies 
at high redshift ($z\sim 1$) are intrinsically brighter than local galaxies.
We postulate a linear evolution
\begin{equation}
  M_{B_j}^*(z) = M^*_{B_j}(z=0) - K_e z \qquad .
\label{eqn:mstarevol}
\end{equation}
and use the 2dF value $M^*_{B_j}(z=0.1) - 5\log_{10}h =-19.79$
(Madgwick et al.~\cite{DSMLF}) to compute the zero point.
By studying the fundamental plane in the rest-frame $B$ band, 
$d\log(M/L)_B /dz$ has been determined to be around $0.4-0.7$
(van Dokkum \& Stanford \cite{vDokSta}, Rusin et al.~\cite{RusinFP}, 
Treu et al.~\cite{Treu}, and references therein). These numbers correspond
to a $K_e$ between 1.0 and 1.75, if $L_*$ galaxies in the past have the same
mass as those of today.
In order to fit the DEEP2 number density and COMBO-17 LF we choose $K_e=1$,
and assume the linear relation of Eq.~\ref{eqn:mstarevol}.

The evolution of $M_*$ is an important assumption that we are putting
into the mock catalogs.  It strongly affects the final $z$-distribution
of galaxies in catalog and we will discuss it further in \S\ref{sec:zdist}.
It also determines which range of intrinsic galaxy luminosities dominate the
DEEP2 sample.  Because of the shape of $L_{\rm cut}/L_*$ shown in
Figure \ref{fig:lumcut} the sample tends to be dominated by galaxies with
$L\simeq 0.5L_*$.

Finally, DEEP2 uses a photo-$z$ cut designed to select only galaxies at
$z>0.7$.  This cut is not a sharp cut in redshift space and the final $z$
distribution turns out to have a peak at $z\sim 0.8$
(see Figure \ref{fig:zhist}) and a foreground tail that extends to quite
low redshift.  To mimic the effects of this cut in our lightcones we applied
a probabilistic foreground cut-off to galaxies with $z<0.8$. 

\section{Basic results from mock catalogs} \label{sec:basic}

By stacking the galaxy boxes in different directions and viewing the
simulations from different positions we produced 12 almost independent
mock catalogs each of which has the size of a DEEP2 field
(or 36 almost independent pointings).

Since the DEEP2 survey is so narrow, with a transverse dimension not much
larger than the non-linear scale, we expect sample variance to be a
significant issue which we can quantify with the mock catalogs.
The simplest statistic is the total counts: the number of galaxies brighter
than the magnitude limit of the survey in each field between redshift
$0.7$ and $1.35$.
The fractional variance in this number is 5.7\%.
For each pointing, which is a third of a field, the fractional variance
is 8.0\%.
We turn now to consider less coarse measures.

\subsection{The cone diagram and the $z$-distribution} \label{sec:zdist}

\begin{figure*}
\begin{center}
\resizebox{7in}{!}{\rotatebox{180}{\includegraphics{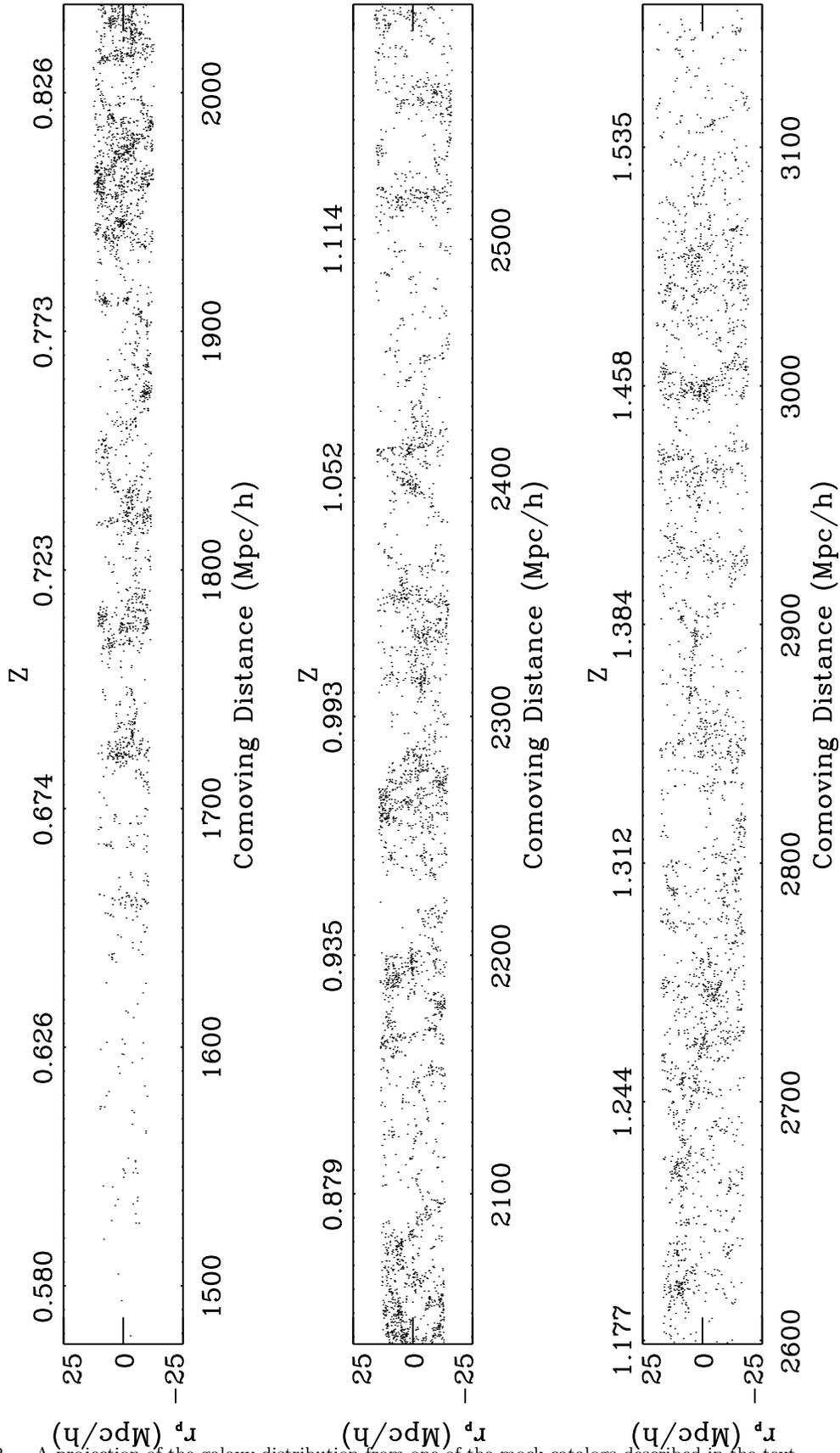}}}
\caption{A projection of the galaxy distribution from one of the mock
catalogs described in the text.}
\label{fig:cone}
\end{center}
\end{figure*}

Figure \ref{fig:cone} shows a cone diagram from one of the mock catalogs
generated from model 4 of Table \ref{tab:nbody}.  Because the DEEP2
geometry is almost a pencil beam, we have broken the redshift direction
up into 3 pieces which we have plotted next to each other.  All of the
galaxies at each redshift have been projected into the slice as a function
of their transverse coordinate.
The large scale structure is visually apparent in the slices, with groups
and clusters separated by voids.
It appears from this figure as if most galaxies lie in large structures
but actually $\sim80\%$ galaxies in the cone are isolated (i.e.~only
one galaxy in their parent halo passes the DEEP2 selection).
Of the dark matter halos with at least 1 galaxy passing the DEEP2 selection,
$\sim90\%$ host only 1 galaxy passing the cut.

\begin{figure}
\begin{center}
\resizebox{3.5in}{!}{\rotatebox{90}{\includegraphics{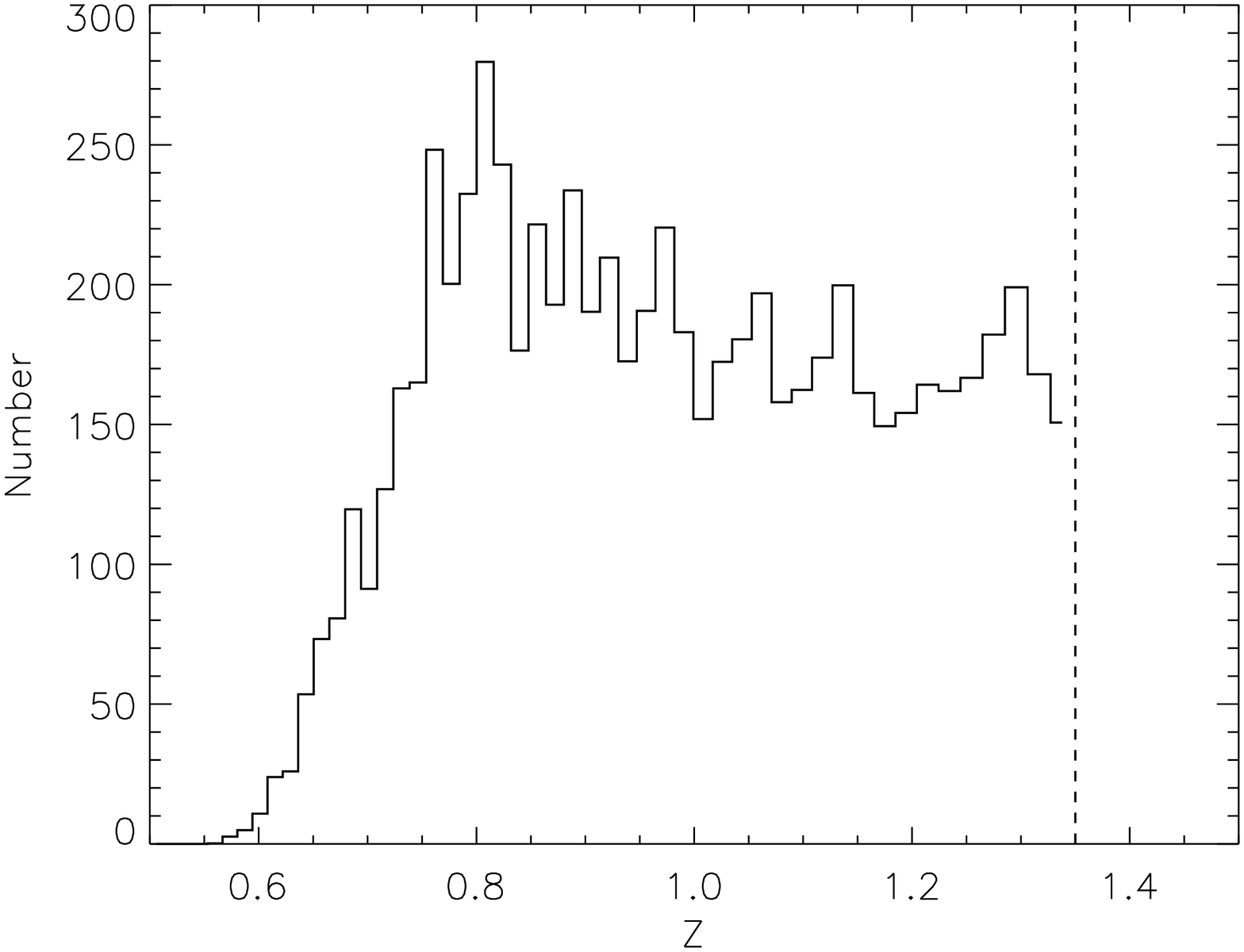}}}
\resizebox{3.5in}{!}{\rotatebox{90}{\includegraphics{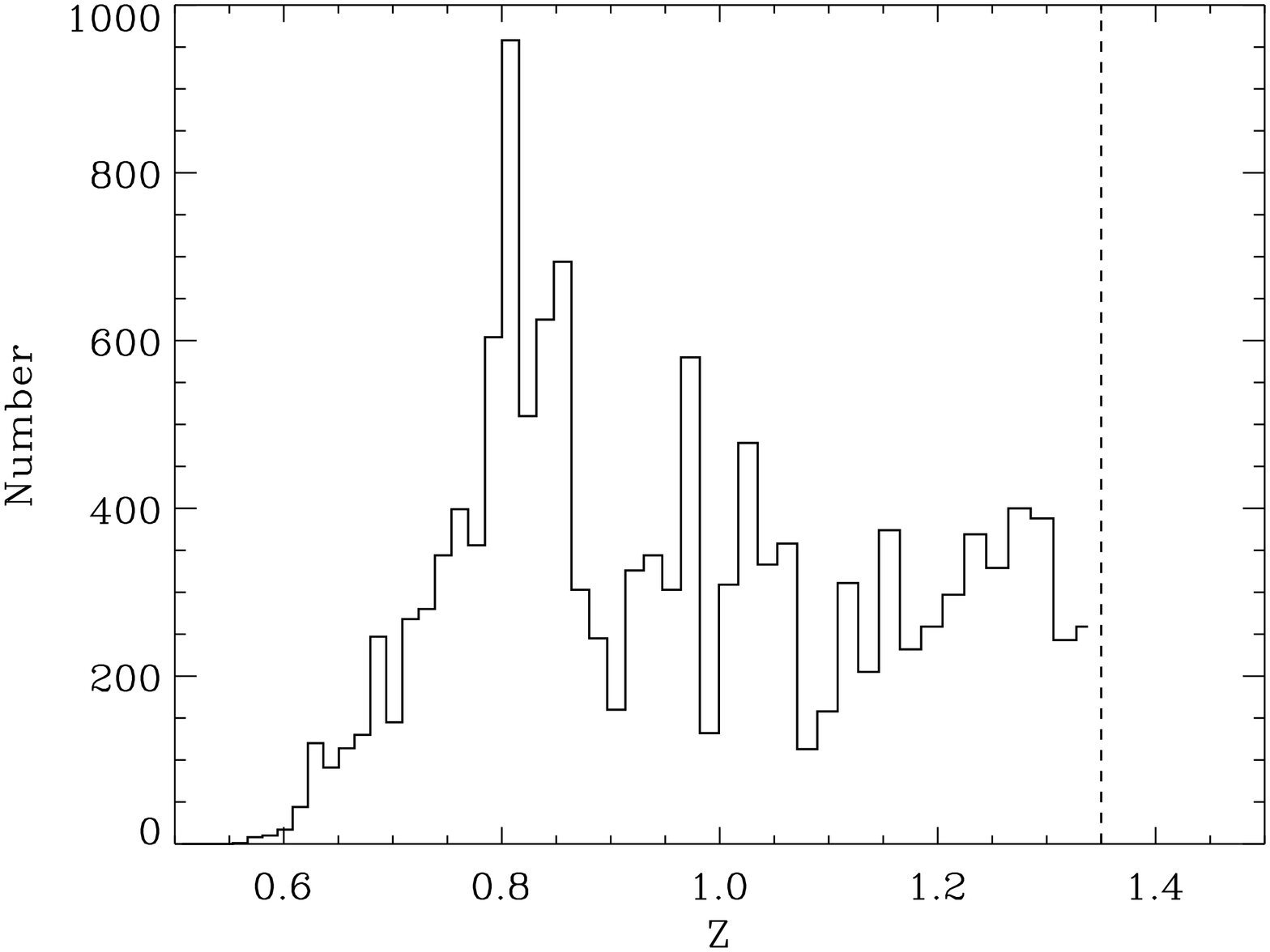}}}
\resizebox{3.5in}{!}{\rotatebox{90}{\includegraphics{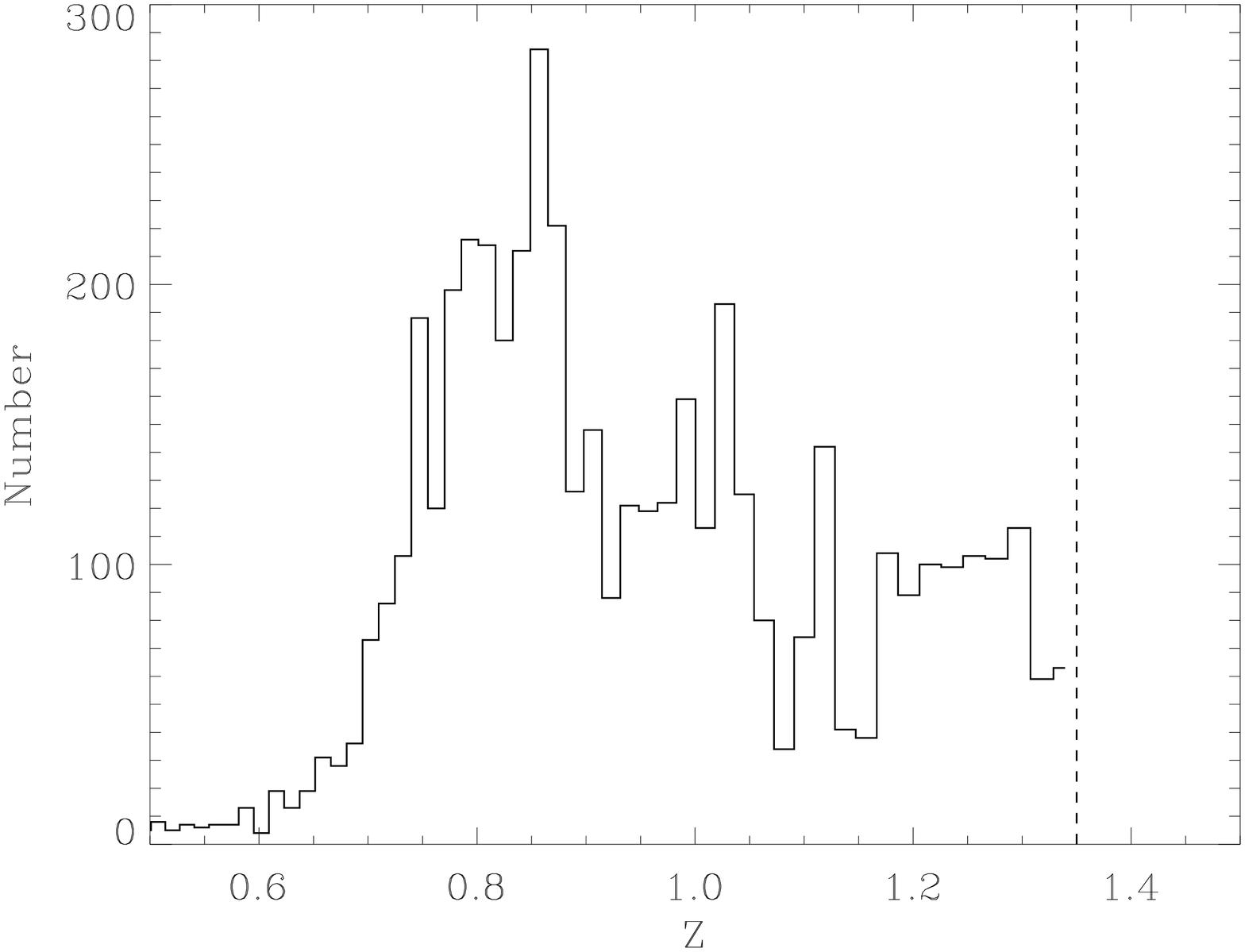}}}
\caption{The upper panel shows the averaged z-distribution over 36 mock
pointings, the middle panel shows that for the sum of two mock pointings 
and the lower
panel shows the observed total z-distribution from three DEEP2 fields
covering an equivalent area of two pointings for comparison.}  
\label{fig:zhist}
\end{center}
\end{figure}

Figure \ref{fig:zhist} compares the $z$ distribution in the mock catalogs
with that in the DEEP2 data.
The observed $z$ distribution is obtained from three separate fields
covering a total of $0.72$ square degrees, which is equivalent to the area
of two `pointings'.
The mocks have similar $z$ distribution to the real data.
The total number of galaxies differs between the mock pointing and the
real data due to various observational effects including the difference
in sky area, the effects of slitmask making and the a non-zero redshift failure
rate.  Also the DEEP2 selection is not complete beyond $z=1.35$ due to the
way slitlets are positioned on the mask, making it difficult to compare to
the data above $z=1.35$.

As we mentioned earlier, the evolution of $L_*$ and the $K$-correction are
important factors affecting the $z$ distribution of galaxies.
We adjust the zero point of these two factors to match the number counts
of galaxies\footnote{When matching the number counts, we assume that among
the galaxies with $z>0.7$, 15\% are at redshifts $z> 1.4$, beyond the
range of DEEP2.} brighter than the magnitude limit of the survey.
Since at present we quote only the luminosity in a single band, this
correction shows up in a combination of the zero points of the $K$-correction
and local $L_*$ measurement.
(In practice, this requires that we increase the apparent magnitude of each
galaxy by $0.1$ to better match the number density.
A better treatment of $K$-corrections requires a knowledge of galaxy types
in the mock catalogue, which we have not implemented yet.)
However, the shape of the $K$-correction and the coefficient of the
$M_*$ evolution, $K_e$, are crucial for the final $z$ distribution.
Figure \ref{fig:lumcut} shows the minimum luminosity as a function of $z$.
Notice that at $z\simeq 0.7$, the minimum luminosity of DEEP2 spectroscopic
targets is around $0.1L_*$, which means for the three fields with a
photo-$z$ cut our mock catalogs probe sufficiently far down the LF.
For the field without a photo-$z$ cut we need to make use of a smaller box,
with higher mass resolution, to extend the light-cones to smaller redshift.

\subsection{Luminosity Function}

\begin{figure}
\begin{center}
\resizebox{3.5in}{!}{\includegraphics{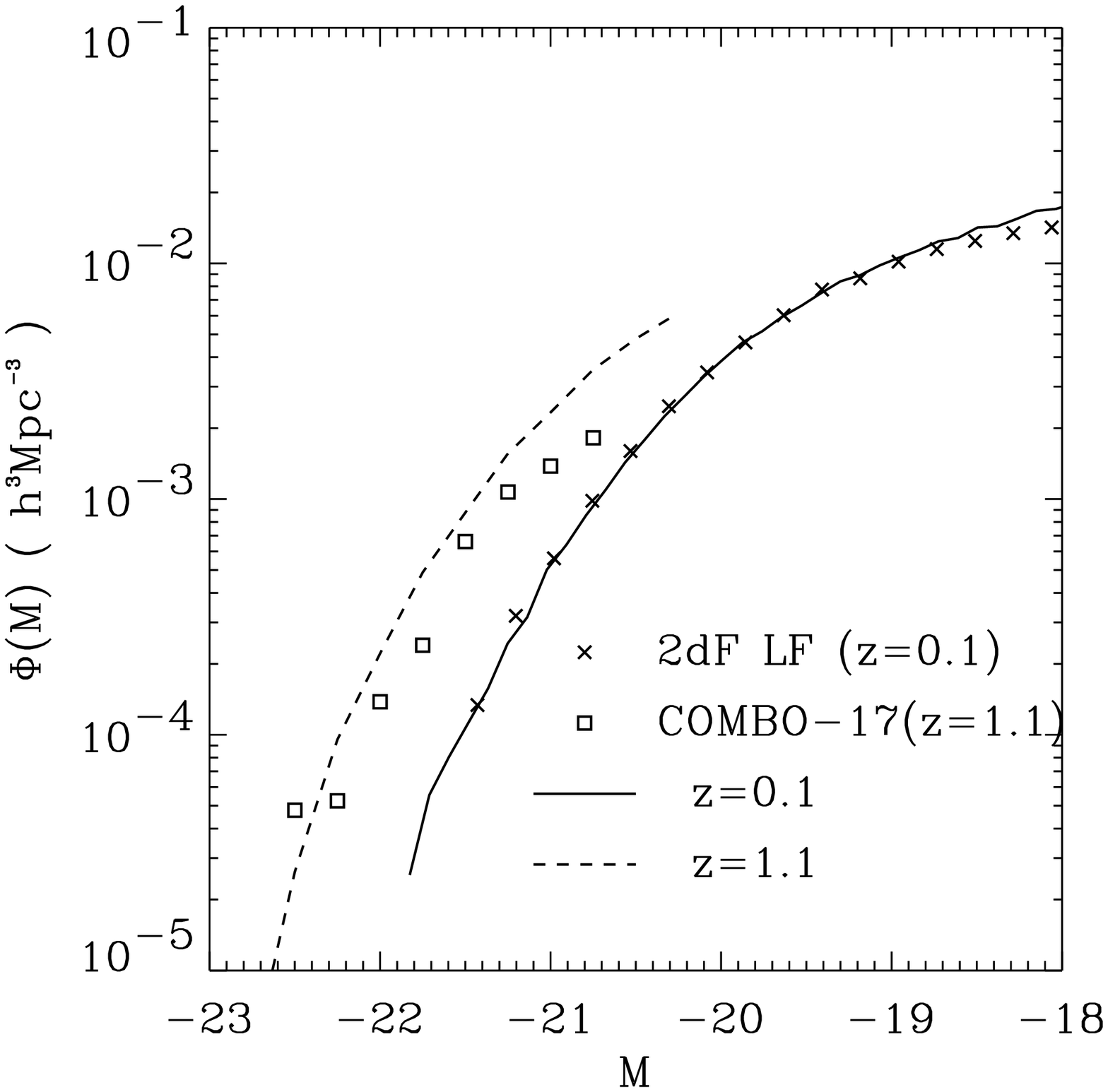}}
\caption{The mean luminosity function of the mock catalogs compared with 
the $B_j$
band COMBO-17 LF for all galaxy SED types in the redshift bin $1.0<z<1.2$.
Also plotted is the comparison between local 2dF $B_j$ band LF and that from
the simulation box, showing the evolution with redshift.}
\label{fig:LFmag}
\end{center}
\end{figure}

Fig \ref{fig:LFmag} compares the LF of the mock with that measured from
COMBO-17 in rest frame $B_j$ band for all SED types.
We measured the LF for galaxies with redshift between 1.0 and 1.2 in each
mock using $1/V_{\rm max}$ method, as was done for COMBO-17.
As is well known, the $1/V_{\rm max}$ method is very sensitive to fluctuations
in number density due to large scale structure in the sample.
We noticed this when comparing the results from 36 different mock catalogs,
some of which were a factor of two off the mean which is plotted in
Figure \ref{fig:LFmag}.
Since the field of view of COMBO-17 is comparable to a DEEP2 pointing, we
regard the difference between the COMBO-17 LF and the mocks as not
statistically significant.
A preliminary LF from DEEP2 is basically consistent with COMBO-17
(Willmer et al.~\cite{ChrisLF}), suggesting our mock catalogs are
an adequate approximation to the DEEP2 data.

\subsection{Projected Correlation Function} \label{sec:wprp}

In addition to the correlation function shown in Figure \ref{fig:Xir},
we computed the projected correlation function $w_p(r_p)$ from the
mock catalogs.  This allows us to compare directly with the measurements
(Coil et al.~\cite{coil}) from early DEEP2 data.
The projected correlation function is almost independent of redshift-space
distortions, since it only uses the transverse spacing between objects.
It is defined as 
\begin{equation}
  w_p(r_p) = 2 \int_0^\infty \xi(r_p,\pi) d\pi \qquad .
\end{equation}
In order to make a direct comparison with the Coil et al.~(\cite{coil})
data we computed $w_p(r_p)$ as described in that paper.
We also divided the mock sample ($0.7 < z < 1.35$) near its median redshift
to make two subsamples: lower-$z$ ($0.7 < z < 0.9$) and higher-$z$
($0.9 < z < 1.35$), performing the $w_p(r_p)$ calculation on each of them. 
First $\xi(r_p,\pi)$ was measured for $|\pi|< 20\,h^{-1}$Mpc then summed
over the $\pi$ direction to give $w_p(r_p)$.  

\begin{figure}
\begin{center}
\resizebox{3.5in}{!}{\includegraphics{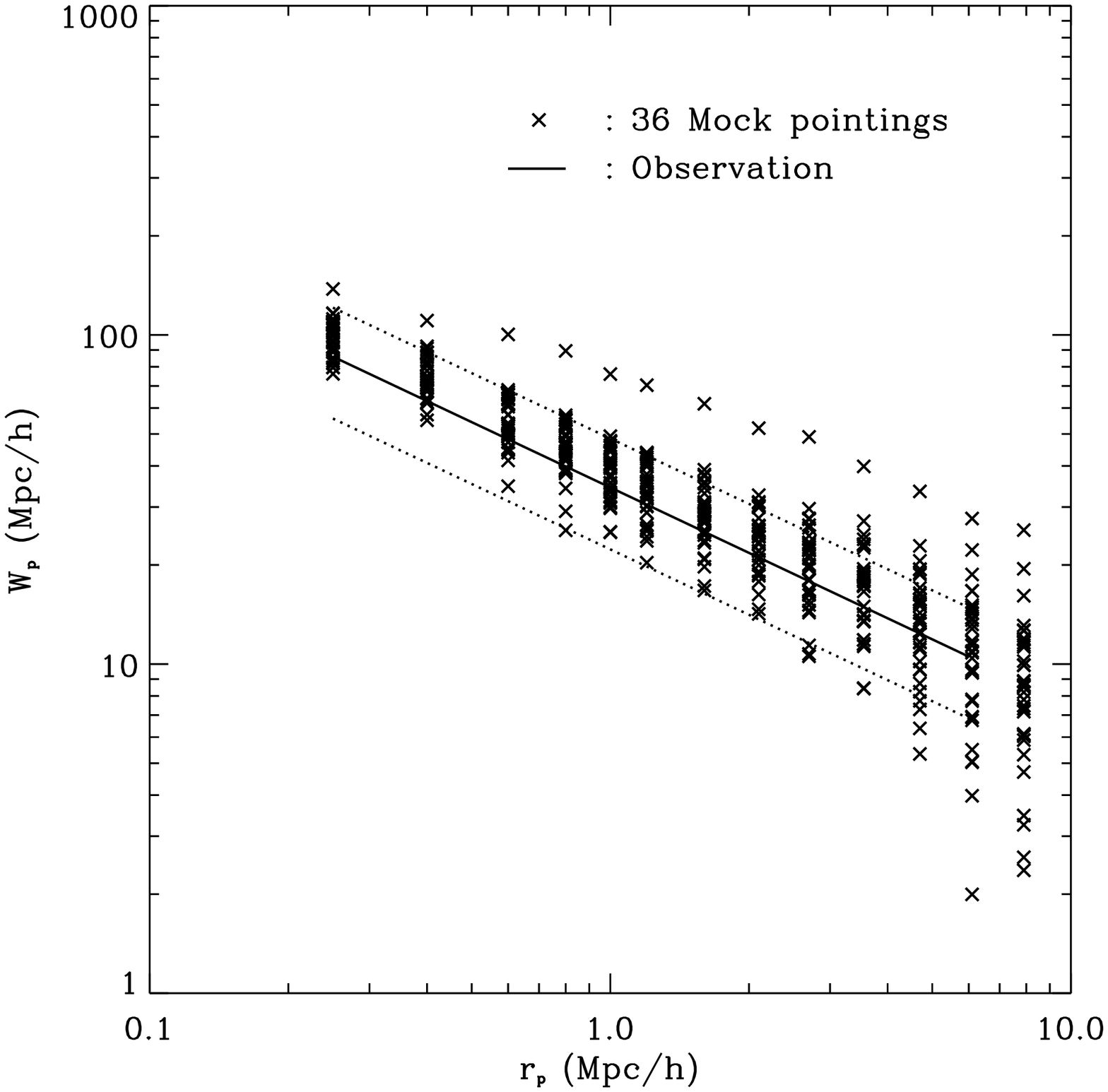}}
\resizebox{3.5in}{!}{\includegraphics{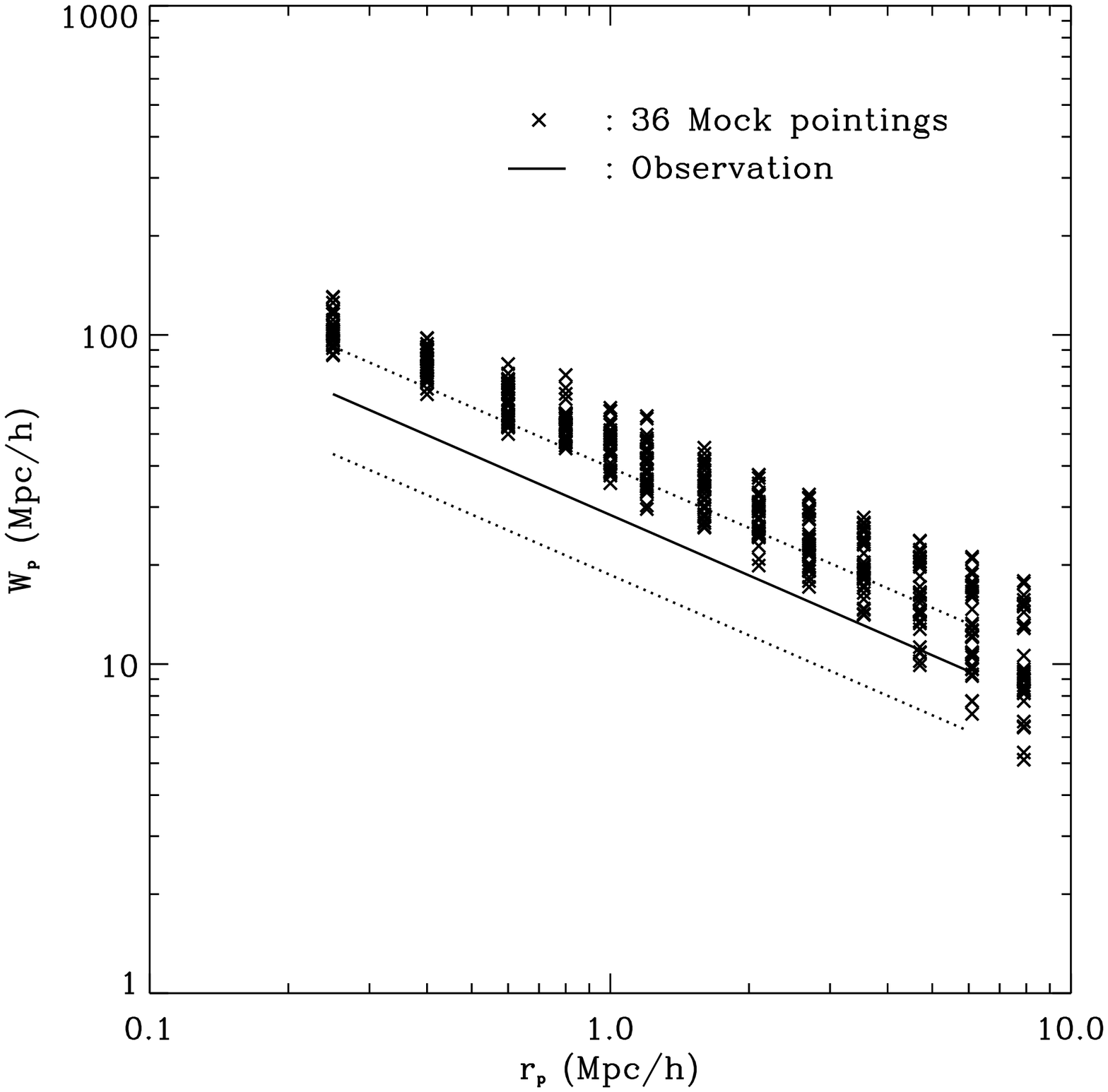}}
\caption{The projected correlation functions from mock catalog compared with
the DEEP2 measurements.  The upper panel shows the lower-$z$ sample and the
lower panel the higher-$z$ sample.  The points (crosses) are measured from
36 mock pointings and the solid lines are the power-law fitting functions from
(Coil et al.~\cite{coil}).  The dotted lines show the $\pm 1\sigma$ error
range, from the error bar on the amplitude assuming a fixed power-law slope.} 
\label{fig:wprp}
\end{center}
\end{figure}

Figure \ref{fig:wprp} shows the results of $w_p(r_p)$ for both subsamples.
The solid lines show the best fit power-law to the DEEP2 data from
Coil et al.~(\cite{coil}), while the dashed lines show the fits for the
$\pm 1\sigma$ range of the amplitude holding the slope fixed.
We cannot include the errors on both the amplitude and slope because they
are highly correlated, and the correlation is not quoted in
Coil et al.~(\cite{coil}).  The error on the slope is smaller than in the
amplitude however, so this range should be a good indication of the current
uncertainty.
The mock points are spread vertically because the sample variance is very
large from field to field.  This is a reflection of the narrowness of one
DEEP2 pointing.
For the lower-$z$ sample, the DEEP2 measurements go almost through the densest
region of the points, showing beautiful agreement between the mocks and the
data for this statistic. We caution that we did not run our mock catalogs 
through the DEEP2 slitmask-making
procedure.  Coil et al.~(\cite{coil}) claim this will only slightly weaken
the correlation strength, and a detailed comparison is beyond the scope of
this paper.

For the higher-$z$ sample, the mocks show slightly stronger clustering
than the data.
For $r_p>1\,h^{-1}$Mpc the data points are covered within the range of
the sample variance of the mocks, but below $1\,h^{-1}$Mpc they are
tantalizingly different.
This discrepancy will be mitigated a little by the complex effects of
mask-making, which we have not included, but likely not enough to bring the
two into agreement.
Further there might be a type-dependent selection bias as we move to higher
redshift, which also needs to be considered.  However this result might be
telling us that the HOD in this redshift range is different from the locally
calibrated HOD that we have used.
This would be very exciting if confirmed, as it indicates that further DEEP2
data can begin to constrain the evolution of the HOD directly through
measurements of clustering at high-$z$.

The correlation function in our mock catalogs is approximately a power-law.
Making a power law assumption we can fit $w_p(r_p)$ in each pointing to
obtain a distribution of spectra indices and amplitudes for $\xi(r)$.
We find that the clustering strength, $r_0$, varies strongly from pointing
to pointing, covering the range $3\,h^{-1}$Mpc to $4.5\,h^{-1}$Mpc.  The
spectral index also varies, and the index and length are clearly correlated.
In the very near future DEEP2 will be able to obtain clustering measurements
over a much wider area of sky than we have at present, which should reduce 
the sample variance. According to our mock catalogs, if correlation functions
are measured over DEEP2 fields instead of pointings, the range of $r_0$ will
shrink by one third.

\section{Limitations and Future development} \label{sec:future}

We hope these mock will be a valuable aid in evaluating the various
selection effects of the survey, testing new algorithm and statistics
which will be used to analyze real data.
But there are several limitations which must be borne in mind when
using these mocks.

\begin{itemize}
\item[(i)] The mocks still present an idealized version of observations.
Real observations have a lot of inevitable difficulties which include
systematic errors in photometry, bright star holes in the field,
CCD defects, incompleteness in the survey caused by weather, failure to
get redshifts for all targeted galaxies, etc.
Some of these difficulties are uncorrectable even with the help of mock
catalog, and this complicates the comparison between mock and reality.
\item[(ii)] We have assumed that our satellite galaxies trace the velocity
field of the dark matter, so no velocity bias (in this sense) is present
in the mock.  This should be borne in mind when comparing galaxy and dark
matter velocity dispersions or the finger of god effect.
\item[(iii)] The halo model and CLF algorithm are a simplification of the
complex physics involved in galaxy formation.  We have chosen a single
specific model for each cosmology to match observational results.
\end{itemize}

To be useful the mock catalogs do not need to be a perfect replica
of the DEEP2 data, they simply have to be a close enough approximation
that they provide a valid test of algorithms and methods.  The required
level of agreement may need to be assessed on a case by case basis.
We view these catalogs as part of a continuing chain of development,
as more information is obtained by DEEP2 the mocks can be further tuned
to increase their fidelity.

We intend to evolve these catalogs as more observational information
becomes available.  In addition to refinements to the CLF parameters and
cosmology we are investigating several extensions to the information
available per galaxy.
First we can include color information in the catalogs.  We are implementing
a scheme for including colors based on local density and luminosity
information, with the model for the color distribution drawn from a
principle component analysis of the data.
Second we are investigating the inclusion of type information similar to
that obtained from the spectral classification scheme of
Madgwick et al.~(\cite{DSMSpec}).
These extensions will be tested against the DEEP2 data as it becomes
available.

\section{Conclusions} \label{sec:conclusions}

We have presented mock catalogs based on high resolution N-body simulations
which are designed to match the low order clustering statistics of the DEEP2
redshift survey.
The mock catalogs are based upon the halo model for clustering and situate
galaxies in their correct cosmological context.
The simulations used are sufficiently large to give many independent
realizations for deep pencil-beam surveys and have sufficiently dense time
sampling to model possible evolution information embedded in the survey.

The mock catalogs provide positions, velocities and luminosities for
galaxies in multiple fields, for a number of cosmologies within the
$\Lambda$CDM class.  The galaxies have a non-trivial scale and luminosity
dependent bias, which is stochastic, but trace the large scale structure
showing voids, filaments, sheets, walls and clusters.
In each case the $z$ distribution, luminosity function, correlation function
and projected correlation function match well the preliminary results from
the DEEP2 survey.

In the process of building these mock catalogs we have noticed several
interesting results.
\begin{itemize}
\item[(i)] The halo model and its extensions (such as the CLF) provide
a powerful tool for building mock catalogs for galaxy redshift surveys.
Galaxies are put into the simulations in a physically motivated way.
This has  two advantages.  First, the HOD and CLF can be constrained from
observations so that it almost guarantees the mocks will mimic reality.
Second, tuning of the model parameters teaches us about galaxy evolution.
The HOD and its extensions provide an ideal meeting ground between
theories of cosmology and galaxy formation.
\item[(ii)] To get the correct $z$ distribution of galaxies is not trivial,
indicating that the observed $z$ distribution contains a lot of information.
Within our formalism, the $z$ distribution involves a $K$-correction,
$L_*$ evolution, halo mass function evolution and CLF evolution.
The first two factors are measurable, the third one depends on cosmology,
and the last one on the details of galaxy formation.
If the $K$-correction and $L_*(z)$ can be well measured, the remaining
degeneracy between cosmology and CLF evolution can be broken by other
measurements, e.g., the fluctuations in the $z$ histogram in a narrow beam
survey, the LF, counts of groups and their richness, the correlation
function etc.
\item[(iii)] Our simple assumption of a HOD which does not evolve from
$z=0$ to $z>1$ provides a reasonable fit to the early DEEP2 data.  This is
a surprising result.  Additionally there may be hints that a non-evolving
HOD will fail to fit improved data at higher redshift, which could provide
valuable information on the formation and evolution of galaxies.
\end{itemize}

We anticipate that these mock catalogs will be crucial in understanding
the data coming from the DEEP2 survey and will help us to understand the
formation and evolution of galaxies.

\acknowledgements
We would like to thank Darren Madgwick for fruitful collaboration
which built the basis of the current work. We would also like to thank
Marc Davis and Jeffrey Newman for numerous enlightening
discussions and Brian Gerke for helpful and detailed comments. We
thank Christopher Willmer for providing the 
$K$-correction. Lastly, we thank the DEEP2 Team for huge effort in this
survey without which the work in this paper would be impossible.
This work was supported in part by the NSF and NASA.
The simulations used here were performed on the IBM-SP at the National
Energy Research Scientific Computing Center.


\begin{thebibliography}{99}
\bibitem[2000]{Benson}
  Benson A. J., Cole S., Frenk C. S., Baugh C. M., Lacey C. G.,
    2000, \mnras, 311, 793
\bibitem[2000]{Blanton}
  Blanton M., 2000, \apj, 544, 63
\bibitem[2002]{BWS}
  Bullock J. S., Wechsler R. H., Somerville R. S., 2002, \mnras, 329, 246
\bibitem[1998]{cole}
  Cole S., Hatton S., Weinberg D.H., Frenk C.S., 1998, \mnras, 300, 945
\bibitem[2001]{CMS}
  Coil A., Davis M., Szapudi I., 2001, \pasp, 113, 1312
\bibitem[2003]{coil}
  Coil A., et al. {\em (the DEEP2 Team)},
    2003, \apj, submitted  [astro-ph/0305586]
\bibitem[2002]{DEEP2}
  Davis M., et al., 2002, Proc. SPIE, 4834, 161 [astro-ph/0209419]
\bibitem[1985]{DEFW}
  Davis M., Efstathiou G., Frenk C.S., White S.D.M., 1985, \apj, 292, 371
\bibitem[1999]{DKCW}
  Diaferio A., Kauffmann G., Colberg J. M., White S. D. M., 1999,
    \mnras, 307, 537
\bibitem[2001]{Feldman}
  Feldman Hume A., Frieman Joshua A., Fry J. N., Scoccimarro R.,
    2001, \prl, 86,1434
\bibitem[2002]{Gray}
  Gray M. E., Taylor A. N., Meisenheimer K., Dye S., Wolf C., Thommes E.,
    2002, \apj, 568, 141
\bibitem[2002]{Hamana}
  Hamana T. et al., 2002, \mnras, 330, 365
\bibitem[2002]{HT}
  Hamilton A.J.S., Tegmark M., 2002, \mnras, 330, 506
\bibitem[2003]{Hatton}
  Hatton S. et al., 2003, \mnras, 343, 75
\bibitem[2002]{Hoekstra}
  Hoekstra H., van Waerbeke L., Gladders M. D., Mellier Y., Yee H. K. C.,
    2002, \apj, 577, 604
\bibitem[1998]{JMB}
  Jing Y. P., Mo H. J., B\"{o}rner G., 1998, \apj, 494, 1
\bibitem[1987]{Kaiser}
  Kaiser N., 1987, \mnras, 227,1
\bibitem[1999]{KCDW}
  Kauffmann G., Colberg J. M., Diaferio A., White S. D. M., 1999,
    \mnras, 303, 188
\bibitem[1997]{KNS}
  Kauffmann G., Nusser A., Steinmetz M., 1997, \mnras, 286, 795
\bibitem[2003]{2MASS}
  Kochanek C., White M., Huchra J., Macri L., Jarrett T.H., Schneider S.E.,
    Mader J., 2003, \apj, 585, 161   [astro-ph/0208168]
\bibitem[2002]{DSMLF}
  Madgwick D.\ S. et al.,  {\em (the 2dFGRS Team)} 2002, \mnras,
    333, 133 [astro-ph/0107197]
\bibitem[2003a]{mad03a}
  Madgwick D.\ S. et al., {\em (the 2dFGRS Team)}
	2003a, \mnras, 344, 847 [astro-ph/0303668]
\bibitem[2003b]{DSMSpec}
  Madgwick D.\ S. et al.,  {\em (the DEEP2 Team)} 2003b,
    \apj, 599, 997 [astroph/0305587]
\bibitem[2001]{Norberg1}
  Norberg P. et al., {\em (the 2dFGRS Team)} 2001, \mnras, 328, 64
\bibitem[2002]{Norberg2}
  Norberg P. et al., {\em (the 2dFGRS Team)} 2002, \mnras, 332, 827
\bibitem[1997]{Peacock}
  Peacock J. A., 1997, \mnras, 284, 885
\bibitem[2000]{PS}
  Peacock J. A., Smith R. E., 2000, \mnras,318,1144
\bibitem[2003]{RusinFP}
  Rusin D. et al., 2003, \apj, 587, 143 [astro-ph/0211229]
\bibitem[2000]{Szapudi}
   Szapudi I., Branchini E., Frenk C. S., Maddox S., Saunders W.,
     2000, \mnras, 318, L45
\bibitem[2002]{ScoShe}
  Scoccimarro R., Sheth R.K., 2002, \mnras, 329, 629S
\bibitem[1996]{LCRS}
  Shectman S.A., Landy S.D., Oemler A., Tucker D.L., Lin H., Kirshner R.P.,
    Schechter P.L., 1996, \apj, 470, 172
\bibitem[1999]{SheTor}
  Sheth R., Tormen G., 1999, \mnras, 308, 119 [astro-ph/9901122]
\bibitem[1999]{SP}
  Somerville R. S., Primack J. R., 1999, \mnras, 310, 1087
\bibitem[1999]{TB}
  Tegmark M., Bromley B.C., 1999, \apj, 518, L69
\bibitem[2002]{Treu}
  Treu T. et al., 2002, \apj, 564, L13
\bibitem[2003]{vdBYM}
  van den Bosch F.C., Yang X., Mo H.J., 2003, \mnras, 340, 771 
       [astro-ph/0210495]
\bibitem[2003]{vDokSta}
  van Dokkum P.G., Stanford S.A., 2003, \apj, 585, 78  [astro-ph/0210643]
\bibitem[1997]{HFW}
  van Haarlem M. P., Frenk C. S., White S. D. M.,1997, \mnras, 287, 817
\bibitem[2001]{HaloMass}
  White M., 2001, A\&A, 367, 27 [astro-ph/0011495]
\bibitem[2002]{TreePM}
  White M., 2002, \apjs, 579, 16 [astro-ph/0207185]
\bibitem[2002]{WhiKoc}
  White M., Kochanek C., 2002, \apj, 574, 24  [astro-ph/0110307]
\bibitem[2004]{ChrisLF}
  Willmer C.  et al. {\em (the DEEP2 Team)}, 2004, in preparation.
\bibitem[2003]{COMBO-17}
  Wolf C. et al., 2003, A\&A, 401, 73 [astro-ph/0208345]
\bibitem[2003]{YMW}
  Yan R., Madgwick D. S., White M., 2003, \apj, 598, 848 [astro-ph/0307248]
\bibitem[2003]{YMvdB}
  Yang X., Mo H-J., van den Bosch F.C., 2003, 
    \mnras, 339, 1057 [astro-ph/0207019]
\bibitem[2003]{YMvdBC}
  Yang X., Mo H-J., van den Bosch F.C., Chu Y., 2003, \mnras, in press
    [astro-ph/0303524]
\bibitem[2001]{Yoshida}
  Yoshida N. et al., 2001, \mnras, 325, 803
\bibitem[2002]{ZJB}
  Zhao D., Jing Y. P., B\"{o}rner G., 2002, \apj, 581, 876
\end{thebibliography}
\end{document}